\documentclass[twocolumn,secnumarabic,amssymb, nobibnotes, aps, prd]{revtex4}
\usepackage{graphicx}

\begin{document}
\title{Comment on "Reversible superconducting-normal phase transition in a magnetic field and the existence of topologically protected loop currents that appear and disappear without Joule heating" by Hiroyasu Koizumi}
\author{A.V. Nikulov}
\affiliation{Institute of Microelectronics Technology and High Purity Materials, Russian Academy of Sciences, 142432 Chernogolovka, Moscow District, RUSSIA.} 

\maketitle 

\narrowtext

\section{Introduction}
The article \cite{Koizumi2020EPL} is based on a not quite correct understanding of the difference between reversible and irreversible thermodynamic processes. The author \cite{Koizumi2020EPL} rightly writes at the beginning of the article that the superconducting-normal transition in the presence of a magnetic field was considered as irreversible before 1933: "{\it At that time it was assumed that the transition in a magnetic field is substantially irreversible, since the superconductor was considered as a perfect conductor (in the sense which was discussed in Chapter II), in which, when the superconductivity is destroyed, the surface current associated with the field are damped with the generation of Joule heat}" \cite{Shoenberg1952}. But all physicists began to consider this transition as a reversible thermodynamic process \cite{Shoenberg1952} after the discovery of the Meissner effect in 1933 \cite{Meissner1933}. 

\section{Reversible thermodynamic processes are described with the help of the free energy}
A transition is reversible if the superconducting state and the normal state have the same free energy $F_{s} = F_{n}$ at the point of the transition $T_{c}$. The free energy used for the description of this transition equals $F = H - ST$ where $H$ is the enthalpy (the sum of the system's internal energy) of normal or superconducting state, $T$ is the absolute temperature, and $S$ is the entropy of normal or superconducting state. All physicists believe after 1933 that the phase transition is observed at the temperature $T = T_{c}$ since $F_{s} > F_{n}$ at $T > T_{c}$ and $F_{s} < F_{n}$ at $T < T_{c}$. According to the point of view of all physicists \cite{Shoenberg1952,Huebener,Tinkham}, the free energy of the superconducting state increases in the magnetic field $H$ by the magnetic field energy equal $F_{m} = \mu _{0}H^{2}/2$ per the unit volume of the bulk superconductor $F_{s}(T,H) = F_{s}(T)+\mu _{0}H^{2}/2$, whereas the energy of the normal state does not change $F_{n}(T,H) = F_{n}(T)$. Therefore the equality 
$$F_{s}(T) + F_{m} = F_{s}(T) + \frac{\mu _{0}H_{c}^{2}}{2} = F_{n}(T) \eqno{(1)}$$
was postulated. $H_{c}(T)$ is the critical field above which superconductivity disappears \cite{Shoenberg1952,Huebener,Tinkham}. The temperature dependence $H_{c}(T)$ is determined by the temperature dependence of the free energy difference $F_{n}(T) - F_{s}(T)$ without the magnetic field energy. The entropy $S = -dF/dT$ changes by jump
$$S_{n} - S_{s} = -\mu _{0}H_{c}\frac{dH_{c}}{dT} \eqno{(2)}$$ 
at the first-order phase transition observed in magnetic field $H_{c}$ \cite{Shoenberg1952}.      

The author \cite{Koizumi2020EPL} assumes that the superconducting-normal phase transition can be reversible if the kinetic energy of the supercurrent is transferable to the magnetic field energy without dissipation. This energy transfer is impossible not only within the standard theory of superconductivity, but also in principle, because of the two reason: 1) both the kinetic energy $F_{k} = V^{-1} \int_{V}dV n_{s}mv^{2}/2 = V^{-1} \int_{V}dV \mu _{0}\lambda _{L}^{2}j^{2}/2 $ and the magnetic field energy increase during the transition from the normal to the superconducting state; 2) the kinetic energy is not considered in textbooks \cite{Shoenberg1952,Huebener,Tinkham} since its value  
$$F_{k} = \mu _{0}H_{c}^{2}\frac{\lambda _{L}}{R} \eqno{(3)}$$  
is much smaller than the magnetic field energy $F_{m} = \mu _{0}H^{2}/2$ of a bulk cylinder with a macroscopic radius $R \gg \lambda_{L}$, where     $\lambda _{L} = (m/\mu _{0}q^{2}n_{s})^{0.5}$ is the London penetration depth; $q = 2e$ is the charge of superconducting pair; $n_{s}$ is the density  of superconducting pairs. If the reversible process would imply $\Delta F_{m} + \Delta F_{k} = 0$, as the author \cite{Koizumi2020EPL} claims, then the superconducting-normal transition could not be reversible in principle. This transition is considered as reversible in all books \cite{Shoenberg1952,Huebener,Tinkham} because of the validity of the equality (1). The enthalpy difference equal $H_{n} - H_{s} = \mu _{0}H_{c}^{2}/2 + (S_{n} - S_{s})T$ at $H = H_{c}(T)$ according to (1) can be found from the experimental results. The entropy difference is positive $S_{n} - S_{s} > 0$ since $dH_{c}/dT < 0$ \cite{Shoenberg1952}. Thus, any theory of superconductivity should explain why the enthalpy decreases on the value $H_{n} - H_{s} = \mu _{0}H_{c}^{2}/2 - T\mu _{0}H_{c}dH_{c}/dT$ in order to describe the normal-superconducting transition as a reversible thermodynamic process. 

The enthalpy decreases due to electron pairing according to the BCS theory \cite{BCS1957}. The author \cite{Koizumi2020EPL} does not consider the enthalpy change because of his wrong assumption that the kinetic energy (3) is transferable to the magnetic field energy (1) during reversible superconducting-normal phase transition. He does not consider also the change of the latent heat $T(S_{n} - S_{s}) = -T\mu _{0}H_{c}dH_{c}/dT$ at this first-order phase transition which is a reversible thermodynamic process as any phase transition. 

The author \cite{Koizumi2020EPL} claims, following Hirsch \cite{Hirsch2020EPL,Hirsch2020ModPhys,Hirsch2020Physica}, that the conventional BCS-London theory of superconductivity cannot explain the normal-superconducting transition because of Joule heating. Hirsch states that "{\it the conventional theory of superconductivity predicts that Joule heat is generated}" \cite{Hirsch2020EPL}. But such direct prediction is absent in the BCS theory \cite{BCS1957}. The presence of such prediction in the theory created within the framework of equilibrium thermodynamics would mean that its authors did not know that Joule heating is an irreversible thermodynamic process. Each physicist must know that the equality of the free energy (1) cannot be valid for an irreversible thermodynamic process.

The author \cite{Koizumi2020EPL} did not quite understand Hirsch's arguments correctly. Hirsch argues that Joule heating occurs at the transition from superconducting to normal state of a metal in a magnetic field \cite{Hirsch2020Physica}. "{\it Joule heating is a non-equilibrium dissipative process that occurs in a normal metal when an electric current flows, in an amount proportional to the metal's resistance}" \cite{Hirsch2020Physica}. The electric current flows during a relaxation time in the normal state with a non-zero resistance after the transition from the superconducting state in the magnetic field $H > H_{c}$. Thus, the superconducting-normal transition in the magnetic field $H_{c}$ cannot be considered as a reversible thermodynamic process according to Hirsch's opinion \cite{Hirsch2020Physica} and in contrast to the opinion of the author \cite{Koizumi2020EPL}. 

\section{Gauge theory of the vector potential}
The author \cite{Koizumi2020EPL} makes a mistake when he uses the London equation \cite{London1950} and the Ginzburg-Landau theory \cite{GL1950}. The distinction, used in \cite{Koizumi2020EPL}, between the vector potential $A$ in the London equation and the ordinary electromagnetic vector potential $A^{em}$ has no sense because of the mathematic equality 
$$\nabla \times \nabla \theta = (\frac{d^{2}\theta}{dydz} - \frac{d^{2}\theta}{dzdy})i_{x} + \cdot \cdot \cdot    \equiv 0 \eqno{(4)}$$ 
which is correct for any function $\theta $. The right-hand side of eq. (16) in \cite{Koizumi2020EPL} is identical to zero because of the mathematical identity (4) rather than because the term with $\nabla \times \nabla \theta$ is often omitted, as the author \cite{Koizumi2020EPL} thinks. 

The term $\nabla \theta$  makes sense in the Ginzburg-Landau equation 
$$j = \frac{q}{m}n_{s}(\hbar \nabla \theta - qA) \eqno{(5)}$$
only if the unambiguity condition of the wave function $\Psi _{GL} =|\Psi _{GL}|\exp{i\theta } = |\Psi _{GL}|\exp{i(\theta + 2\pi n)}$ is used \cite{Huebener}. According to this condition $\oint_{l}dl \nabla \theta = n2\pi $ and the quantization of the density of superconducting current 
$$\mu _{0}\oint_{l}dl \lambda _{L}^{2} j  + \Phi = n\Phi_{0}  \eqno{(6)}$$
along a closed path $l$ is deduced. Here $\oint_{l}dl A = \Phi $ is the magnetic flux inside $l$; $n$ is an integer number and $\Phi _{0} = 2\pi \hbar /q$ is the flux quantum. The quantization of the magnetic flux $\Phi = n\Phi_{0}$ is observed \cite{fluxquan1961} when $\oint_{l}dl j = 0$ along $l$. The Meissner effect $\Phi = 0$ is observed when the wave function has no singularity inside $l$ and therefore $n = 0$   

\acknowledgments
This work was made in the framework of State Task No 075-00355-21-00.

\end{document}